\documentclass[iicol]{sn-jnl}

\usepackage{amssymb}
\usepackage{latexsym}
\usepackage{graphicx}
\usepackage{amsfonts,amssymb,amsmath}
\usepackage{hyperref}
\usepackage{adjustbox}
\usepackage{array}
\usepackage[utf8]{inputenc}
\usepackage[english]{babel}
\usepackage{xcolor}
\usepackage{caption} 
\usepackage{multirow,booktabs, makecell}
\usepackage{subfigure}

\jyear{2021}%

\theoremstyle{thmstyleone}%
\theoremstyle{thmstyletwo}%

\theoremstyle{thmstylethree}%

   \usepackage{setspace}

\begin{document}

\title[Article Title]{Revisiting Table Detection Datasets for Visually Rich Documents}

\author[1]{\fnm{Bin} \sur{Xiao}}\email{bxiao103@uottawa.ca}

\author[1]{\fnm{Murat} \sur{Simsek}}\email{murat.simsek@uottawa.ca}

\author*[1]{\fnm{Burak} \sur{Kantarci}}\email{burak.kantarci@uottawa.ca}

\author[2]{\fnm{Ala Abu} \sur{Alkheir}}\email{ala\_abualkheir@lytica.com}

\affil[1]{\orgname{School of Electrical Engineering and Computer Science}, \orgaddress{\street{University of Ottawa}, \city{Ottawa}, \postcode{K1N 6N5}, \state{ON}, \country{Canada}}}

\affil[2]{\orgname{Lytica}, \orgaddress{\street{555 Legget Dr}, \city{Ottawa}, \postcode{K2K 2X3}, \state{ON}, \country{Canada}}}

\abstract{Table Detection has become a fundamental task for visually rich document understanding with the surging number of electronic documents. However, popular public datasets widely used in related studies have inherent limitations, including noisy and inconsistent samples, limited training samples, and limited data sources. These limitations make these datasets unreliable to evaluate the model performance and cannot reflect the actual capacity of models. Therefore, this study revisits some open datasets with high-quality annotations, identifies and cleans the noise, and aligns the annotation definitions of these datasets to merge a larger dataset, termed Open-Tables. Moreover, to enrich the data sources, we propose a new ICT-TD dataset using the PDF files of Information and Communication Technologies (ICT) commodities, a different domain containing unique samples that hardly appear in open datasets. To ensure the label quality of the dataset, we annotated the dataset manually following the guidance of a domain expert. The proposed dataset is challenging and can be a sample of actual cases in the business context. We built strong baselines using various state-of-the-art object detection models. Our experimental results show that the domain differences among existing open datasets are minor despite having different data sources. Our proposed Open-Tables and ICT-TD can provide a more reliable evaluation for models because of their high quality and consistent annotations. Besides, they are more suitable for cross-domain settings. Our experimental results show that in the cross-domain setting, benchmark models trained with cleaned Open-Tables dataset can achieve 0.6\%-2.6\% higher weighted average F1 than the corresponding ones trained with the noisy version of Open-Tables, demonstrating the reliability of the proposed datasets. The datasets are public available at \url{http://ieee-dataport.org/documents/table-detection-dataset-visually-rich-documents}. }




\maketitle

\section{Introduction}
\label{sec:introduction}
\begin{figure*}[htp]
\begin{center}
  \includegraphics[width=0.6\textwidth]{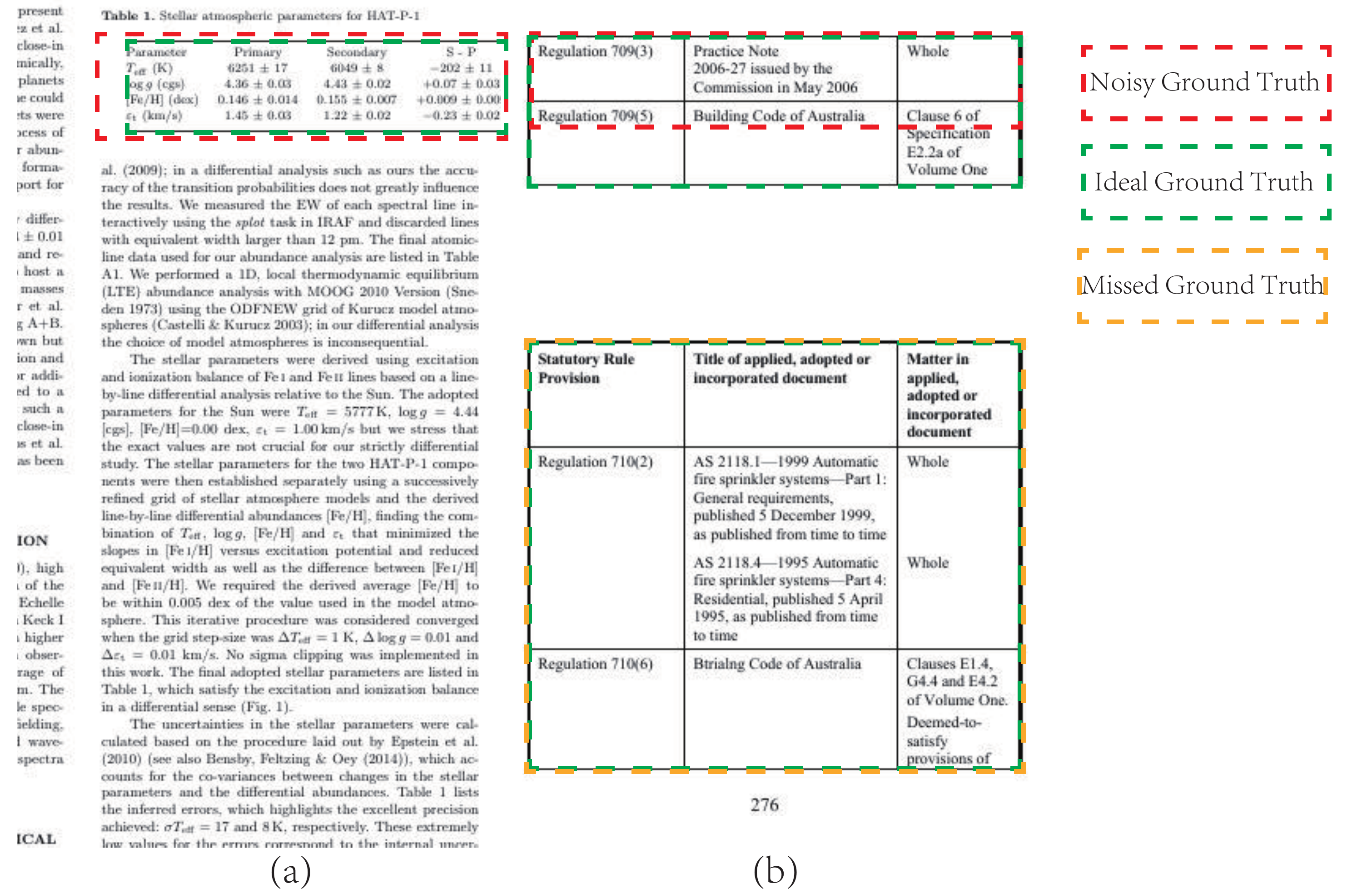}
  \caption{Two noisy samples from the TableBank~\cite{li2020tablebank} testing set. Red bounding boxes are the ground truth boxes provided by the dataset. In Figure (a), the bounding box is larger than the ideal bounding box, making the evaluation unreliable when the IoU threshold is high. Figure (b) shows a sample whose bounding box needs to be larger to cover the whole table. Besides, the table at the bottom of Figure (b) is not annotated. It is worth mentioning the images' low resolution is caused by the images provided by the dataset.}
  \label{fig:noisy_samples}
\end{center}
\end{figure*}

Tables or tabular data have been widely used in electronic documents to summarize critical information so that the information can be presented efficiently to human readers. However, electronic documents, such as Portable Document Format (PDF) files, cannot provide enough meta-data to describe the location and the structure of these tables, meaning that these tables are unstructured and cannot be quickly processed and interpreted automatically. With the surging amount of electronic documents, Table Detection (TD) becomes a fundamental task for downstream document understanding tasks, such as Key Information Extraction and Table Structure Recognition \cite{akkaya2022cropped}. With the development of deep learning, transforming electronic documents into visually rich document images and formulating the problem as an object detection problem became the dominant solutions. There have been some public datasets for the TD problem, such as ICDAR2013~\cite{gobel2013icdar}, ICDAR2017~\cite{gao2017icdar2017}, ICDAR2019~\cite{gao2019icdar} and TableBank~\cite{li2020tablebank}. 
Some of these datasets are manually labeled, meaning the annotations are more reliable and consistent, but the number of training sample in these datasets are usually limited. Besides, the annotation definitions across these datasets are often different, so we cannot simply merge these datasets to form larger datasets. In contrast, datasets such as TableBank~\cite{li2020tablebank} and PubLayNet~\cite{zhong2019publaynet} are annotated by parsing meta-data of electronic documents, making these annotations noisy and inconsistent, even though these datasets are much larger. Figure~\ref{fig:noisy_samples} shows two samples from the TableBank test set. One typical issue of these meta-data generated datasets is that the bounding box can be larger than an ideal bounding box, as shown in Figure~\ref{fig:noisy_samples} (a), which can make the evaluation unreliable when the Intersection over Union (IoU) threshold is high. Another issue is that some tables are missing, or their bounding boxes are not large enough to cover the whole table, as shown in Figure~\ref{fig:noisy_samples} (b). The quality of a table detection set is critical for the TD problem because a successful TD application should avoid losing information presented in the tables. The noisy labels in the test set can influence the model evaluation, especially for widely used evaluation metrics threshold by IoU scores. It is worth mentioning that even though manually annotated datasets have a higher quality of annotations, there are still many noisy samples in both their training and testing sets. Therefore, in this study, we revisit several well-annotated datasets, including ICDAR2013, ICDAR2017, ICDAR2019, Marmot, and TNCR, align the labeling definition of these datasets, clean the noisy samples and merge them to form a larger dataset, termed Open-Tables. The new Open-Tables dataset can minimize the side effects of noisy samples on the model evaluation and provide more reliable results. We include more details regarding the Open-Tables dataset in section~\ref{sec:open_tables}.

\begin{figure*}[htp]
\begin{center}
  \includegraphics[width=0.6\textwidth]{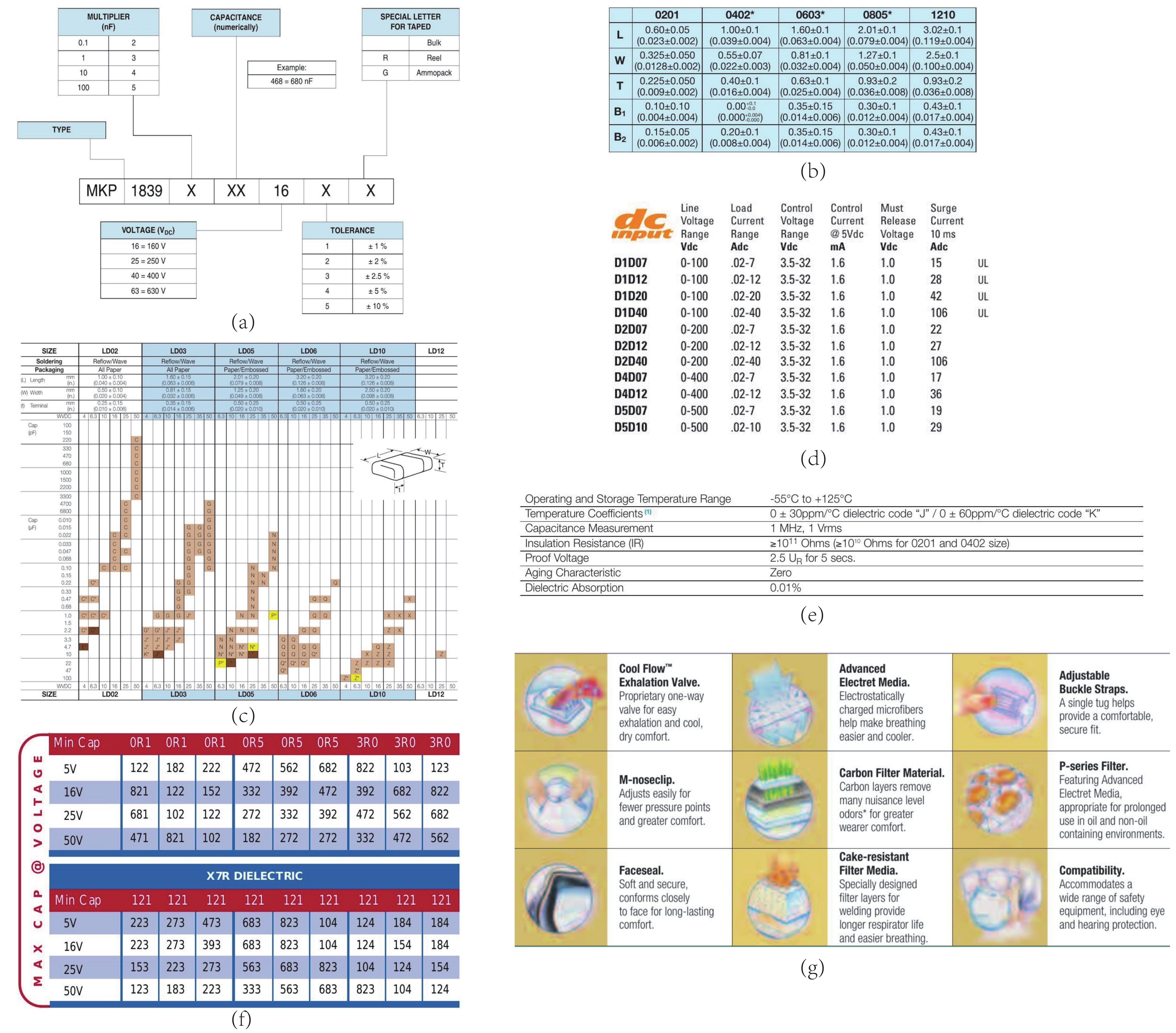}
  \caption{Table examples from the proposed ICT-TD dataset.}
  \label{fig:table_samples}
\end{center}
\end{figure*}

Besides the issues of noisy labels, the data sources of open datasets are limited, primarily from academic publications or public governmental documents. The limited data sources make the intra-class and inter-class variance of these documents small because these documents have to be written following a series of writing principles. In other words, detection models can easily achieve a promising performance on these datasets, which also cannot reflect the complexity of real enterprise applications. Therefore, we propose a new TD dataset using datasheets from the Information and Communication Technologies (ICT) domain in this study. It is a more challenging dataset because of the domain-specific samples, layout, and appearance variances. Figure~\ref{fig:table_samples} shows some samples from our proposed dataset, which can hardly be found in the public datasets. For example, Figure~\ref{fig:table_samples} (a) is a table containing several sub-tables, and Figure~\ref{fig:table_samples} (7) is a table containing figures as the content of some table cells. We list a few examples from the proposed dataset in Figure~\ref{fig:table_samples}, making it challenging and different from other open datasets. We will include more details regarding the proposed dataset in section~\ref{sec:ict_td_dataset}.

We also build strong baselines using the state-of-the-art object detection models including TableDet~\cite{fernandes2022tabledet}, DiffusionDet~\cite{chen2022diffusiondet}, Deformable-DETR~\cite{zhu2020deformable} and SparseR-CNN~\cite{sun2021sparse} for the Open-Tables and ICT-TD datasets. Because of the obvious domain difference between the Open-Tables and ICT-TD datasets, baselines in the cross-domain settings are also built to evaluate the model generalization capacity further. The Open-Tables and ICT-TD datasets can accelerate the study of typical and cross-domain TD problems and minimize the side effects of noisy samples on the model evaluation.

To sum up, the contribution of this article are three-fold:
\begin{enumerate}

\item This study revisits several open TD datasets, align their annotations, clean the noisy labels and merge them to form a new dataset termed with Open-Tables. Open-tables dataset can provide more reliable model evaluation minimizing the side effects of noisy labels.

\item A new manually annotated dataset, termed with ICT-TD, is proposed using the PDF files of ICT commodities containing many domain-specific training samples. The ICT-TD dataset provides a new data source and contains many unique samples that can hardly be found in other open datasets.

\item In addition to a variety of strong baselines using different types of state-of-the-art object detection methods, this study also provides benchmarks in the cross-domain setting.

\end{enumerate}

The rest of this paper is organized as follows: Section~\ref{sec:related_work} discusses related studies, including related datasets, studies in Object Detection Models and Table Detection models. Section~\ref{sec:proposed_method} introduces our proposed Open-tables and ICT-TD datasets, the formal problem definition, and the baseline models. Section~\ref{sec:experiments}   presents the experimental settings and results and builds the baselines for our proposed dataset. At last, we draw our conclusion and possible directions in Section~\ref{sec:conclusion}.

\section{Related Work}
\label{sec:related_work}
As discussed in section~\ref{sec:introduction}, the TD problem is usually formulated as an object detection problem. This section discusses related datasets, popular object detection models, and table detection models.

\subsection{Related Datasets}
\label{sec:related_datasets}
There have been many public datasets that can be used for the TD problem. Some of these datasets are created only for the TD problem, such as ICDAR2013~\cite{gobel2013icdar}, ICDAR2017~\cite{gao2017icdar2017}, ICDAR2019~\cite{gao2019icdar} and TNCR~\cite{abdallah2022tncr}. The ICDAR2013 dataset is a widely used benchmark in many studies, which contains 238 images collected from Public Governmental Documents. Some images in the ICDAR2013 dataset do not contain any tales, resulting in 150 tables. Since this dataset is relatively small, many studies have achieved 100\% F1-score, following the evaluation metric setting of ICDAR2013 competition~\cite{gobel2013icdar} whose IoU threshold is 0.5. ICDAR2017 and ICDAR2019 datasets are the other two used in the corresponding competitions. The ICDAR2017 dataset consists of a training set with 1600 images and a testing set with 817 images. The documents in the ICDAR2017 are collected from CiteSeer, which is a source of academic publications. ICDAR2019 has two separate collections, including a set of archival document images and a set of modern document images. In this study, we only consider the modern document set of the ICDAR2019 dataset, which contains 600 images for training and 240 for testing. TNCR~\cite{abdallah2022tncr} also collects the Public Governmental Documents and further defines five types of tables, including Full-lined, No-lined, Merged-cells, Partial lined, and Partial lined merged cells. TNCR contains 4634, 987, and 1000 images for training, testing, and validation images, respectively. IIIT-AR-13K~\cite{mondal2020iiit} is another dataset defining five types of document objects, including Table, Figure, Natural Image, Logo, and Signature. IIIT-AR-13K dataset uses business annual reports as the data source, consisting of 9000 training images, 2000 testing images, and 2000 validation images. 

Some datasets are generated by parsing the meta-data of Microsoft Word files or latex sources, often much larger than manually annotated datasets. TableBank~\cite{li2020tablebank} is a typical automatically generated dataset containing latex and Microsoft Word generated samples. The Word files are collected from the internet, and the latex source codes are from open academic publications. Similarly, PubLayNet~\cite{zhong2019publaynet} dataset is generated by parsing the XML format documents of academic publications. PubLayNet defines five types of document components, including Text, Title, List, Table, and Figure. We summarize the statistics of these open datasets in Table~\ref{table:dataset_statistics}. 

As discussed in Section~\ref{sec:introduction}, the open datasets have some inherent limitations. Firstly, the annotation definitions across these datasets can differ, especially for the ambiguous samples. Second, the data sources of these datasets are limited, mainly from academic publications and open governmental documents. Even though there are other data sources, such as business annual reports, the samples from these sources are usually simple and similar to those from other open datasets. Third, these open datasets can be noisy, especially for these automatically generated datasets, which can make the model evaluation using these datasets are not reliable. Besides, the data sources of these datasets are very limited, mainly from academic publications and open governmental documents. These limitations make the models trained with datasets hardly have good generalization ability to the different domains. Besides, the noise samples in the training set can hinder the model performance, and the noise samples in the testing set can make the model evaluation unreliable. Therefore, to alleviate these limitations, this study proposes the Open-Tables and ICT-TD datasets, which have consistent annotation and less noise. Besides, this study also builds the benchmarks for the cross-domain setting, which is challenging but valuable for TD applications.

\subsection{Object Detection Models}
\label{sec:object_detection_models}
Object Detection problem is a popular topic in Computer Vision and has been widely discussed in recent years. We can categorize deep learning based object detection models into three types: one-stage, two-stage, and transformer-based. One-stage and two-stage models rely heavily on Convolution Neural Network(CNN) and follow a region detection and object classification pipeline. The main difference between these model types is whether these two tasks can be solved in a single deep neural network. Taking YOLO~\cite{redmon2016you} as an example of the one-stage model, the first step of YOLO is dividing the input image into N * N grid cells. Each grid cell is used to predict a confidence score containing a target object and a conditional class probability for target classes. All these steps are finished within a single model. In contrast, two-stage models, such as Faster-RCNN~\cite{ren2015faster}, usually employ a Region Proposal Network (RPN) to generate a series of region proposals. These region proposals are fed into an ROIHead to classify the object class and refine the bounding boxes by a regression task. In two-stage models, region proposals are generated by a separate network, making them have two stages. One-stage models are usually faster than two-stage models regarding the inference time, while two-stage models usually can outperform one-stage models regarding the detection performance. There have been many studies ~\cite{liu2016ssd, redmon2016you, li2017fssd, ning2017inception, shafiee2017fast, redmon2018yolov3, bochkovskiy2020yolov4, sun2021sparse, he2017mask} following these typical design of one-stage and two-stage model, refining some parts of the model, such as region proposal methods, backbone networks, adding other sub-tasks to the multi-task architecture. With the development of the self-attention mechanism~\cite{vaswani2017attention}, transformer architecture is also adapted to the object detection problem. DETR~\cite{zhu2020deformable} is the first study bridging Transformer architecture and object detection problem. In DETR, there is an embedding network implemented by a popular CNN network to generate the image features, and then the features are fed into an Encoder-Decoder architecture implemented by Transformers, and the problem is formulated as a bipartite matching problem. DETR can achieve state-of-the-art performance, but it converges very slowly. Therefore, there have been some studies, such as Deformable DETR~\cite{zhu2020deformable}, Conditional DETR~\cite{chu2021conditional}, Dynamic DETR~\cite{dai2021dynamic} and Fast DETR~\cite{gao2021fast}, trying to improve the performance and speed up DETR by introducing different positional encoding method and refining the attention modules. Diffusion models are widely used for image generation problems and were first adopted to the Object Detection problem by DiffusionDet~\cite{chen2022diffusiondet}. Similar to typical diffusion models, DiffusionDet~\cite{chen2022diffusiondet} also consists of forward and reverse diffusion processes. Still, the bounding boxes are the target of these two processes, noising and denoising. 

\subsection{Table Detection}
\label{sec:table_detection}
Table Detection is a fundamental step for downstream tasks such as key information extraction and visually rich document understanding. Typically, The table detection problem is formulated as an object detection. Considering the difference between objects in natural images and tables in the image documents, YOLOv3-TD~\cite{huang2019yolo} proposes an anchor optimization strategy and two post-processing methods to adjust the detection method. The authors of YOLOv3-TD observe that the width of a table is usually larger than its height unless the table is very big. Based on this observation, they propose a K-means based method to optimize anchors and obtain more “horizontal” anchors. Besides, they also erase the white space margin from predicted regions and filter the noisy page objects as the post-processing methods to improve the model performance further. CascadeTabNet~\cite{prasad2020cascadetabnet} is a typical two-stage model that is applied to the Table Detection problem. CascadeTabNet is based on Cascade Mask R-CNN~\cite{cai2019cascade} with HRNet~\cite{wang2020deep} as the backbone network. In addition, CascadeTabNet also employs an image augmentation method, which can thicken the text regions and reduce the regions of blank space, and utilizes a transfer learning approach to train the model iteratively. Similar to one-stage and two-stage models, transformer-based approach, such as DETR~\cite{carion2020end} is also discussed in some studies~\cite{smock2022pubtables} for the TD problem. There are also many other studies discussing the TD problem, such as TableDet~\cite{fernandes2022tabledet}, DeCNT~\cite{siddiqui2018decnt}, DeepDeSRT~\cite{schreiber2017deepdesrt}, TableNet~\cite{paliwal2019tablenet}, and most of these studies follow the object detection formulation and utilize different types of object detection models that are mentioned above. It is worth noting that studies~\cite{kara2020holistic, jiang2020high} also discuss the TD problem for the ICT domain, but the dataset used in these two studies is smaller than the proposed ICT-TD dataset in this study. 

\section{Proposed Dataset}
\label{sec:proposed_method}
As aforementioned, existing open-source table detection datasets usually need to be more complex to reflect the complexity and difficulty of business scenarios. This section provides the details of the proposed Open-Tables and ICT-TD datasets.

\subsection{Open-Tables Dataset}
\label{sec:open_tables}
In this section, we discuss the noise cleaning and the annotation alignment for the ICDAR2013~\cite{gobel2013icdar}, ICDAR2017~\cite{gao2017icdar2017}, ICDAR2019~\cite{gao2019icdar}, Marmot~\cite{fang2012dataset}, and TNCR~\cite{abdallah2022tncr} datasets to create the Open-Tables dataset. 

As discussed in section~\ref{sec:related_datasets}, ICDAR2019 contains archival and modern documents. We only use the modern documents in this study. Since there are five find-grained types of tables in the TNCR dataset, we transform all these annotations into a single type, namely tables. Even though the annotation quality of the datasets used here is relatively higher, many samples still have noisy annotations that have the issues shown in Figure~\ref{fig:noisy_samples}. These samples in the test set can influence the model evaluation, and noisy samples in the training set can degrade the model performance. Therefore, we first corrected these noisy samples.

\begin{figure*}[htp]
\begin{center}
  \includegraphics[width=0.6\textwidth]{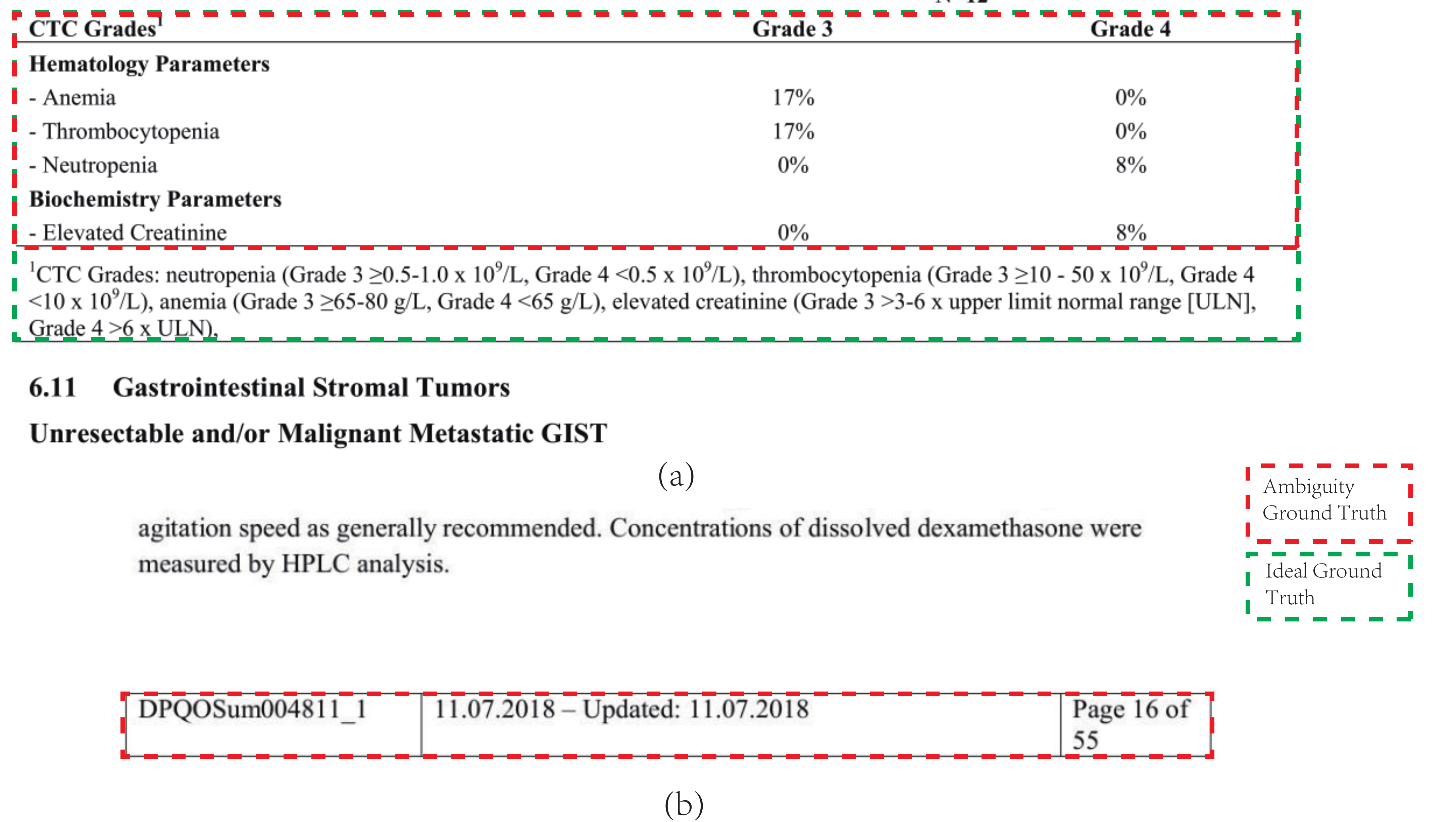}
  \caption{Two ambiguous samples. Figure (a) shows two alternative annotations of a table. The green bounding box in Figure (a) is the ground truth provided by the TNCR dataset, which excludes the table explanation part. The red bounding boxes in Figure (b) are defined ground truth but are not tables in other datasets.}
  \label{fig:ambiguity_samples}
\end{center}
\end{figure*}

Besides the noisy annotations, as discussed in section~\ref{sec:introduction} and ~\ref{sec:related_datasets}, the table definition across these open datasets can differ. This issue is caused by ambiguous samples. Figure~\ref{fig:ambiguity_samples} shows two ambiguous samples from the TNCR dataset. The first ambiguous sample shows two alternative bounding boxes that can cause inconsistent annotation issues. The ground truth (green box) provided by the TNCR dataset excludes the explanation part of the table. The second ambiguous sample is labeled as table but is unnecessary in other datasets because it can also be defined as a document footer. To address these ambiguous samples, we define the following rules to align the datasets. First, we use the table lines as the priority, meaning we include all the content bounded by the table lines. However, when table lines do not bind a table explanation part, it should not be defined as part of the table. Second, a table should at least have two lines and two columns. Following these two rules, we should include the explanation part of the first sample and define the second sample as none-table, as shown in Figure~\ref{fig:ambiguity_samples}.

\subsection{ICT-TD Dataset}
\label{sec:ict_td_dataset}
In this section, we discuss the data collection and data pre-processing for the proposed ICT-TD dataset. We collect 175,682 PDF documents for 370 different ICT commodities. Since each PDF file may have more than one page, we transform each page into an image with a resolution of 200 DPI, resulting in 3,581,805 images. We employ a random sampling method to select 5,000 samples containing tables from these images and manually annotate the bounding boxes of all the tables in the images. We summarize the statistics of the ICT-TD dataset and some public datasets in Table~\ref{table:dataset_statistics} for comparison purposes. ICDAR2013~\cite{gobel2013icdar} is a small dataset without providing a training set. ICDAR2017~\cite{gao2017icdar2017}, ICDAR2019~\cite{gao2019icdar}, Marmot~\cite{fang2012dataset}, and TableBank~\cite{li2020tablebank} are all using academic publications or public governmental documents as the data sources, making these datasets cannot reflect the complexity of real enterprise cases and hard to be adapted to the ICT domain. 

\begin{figure}[htp]
\begin{center}
  \includegraphics[width=\columnwidth]{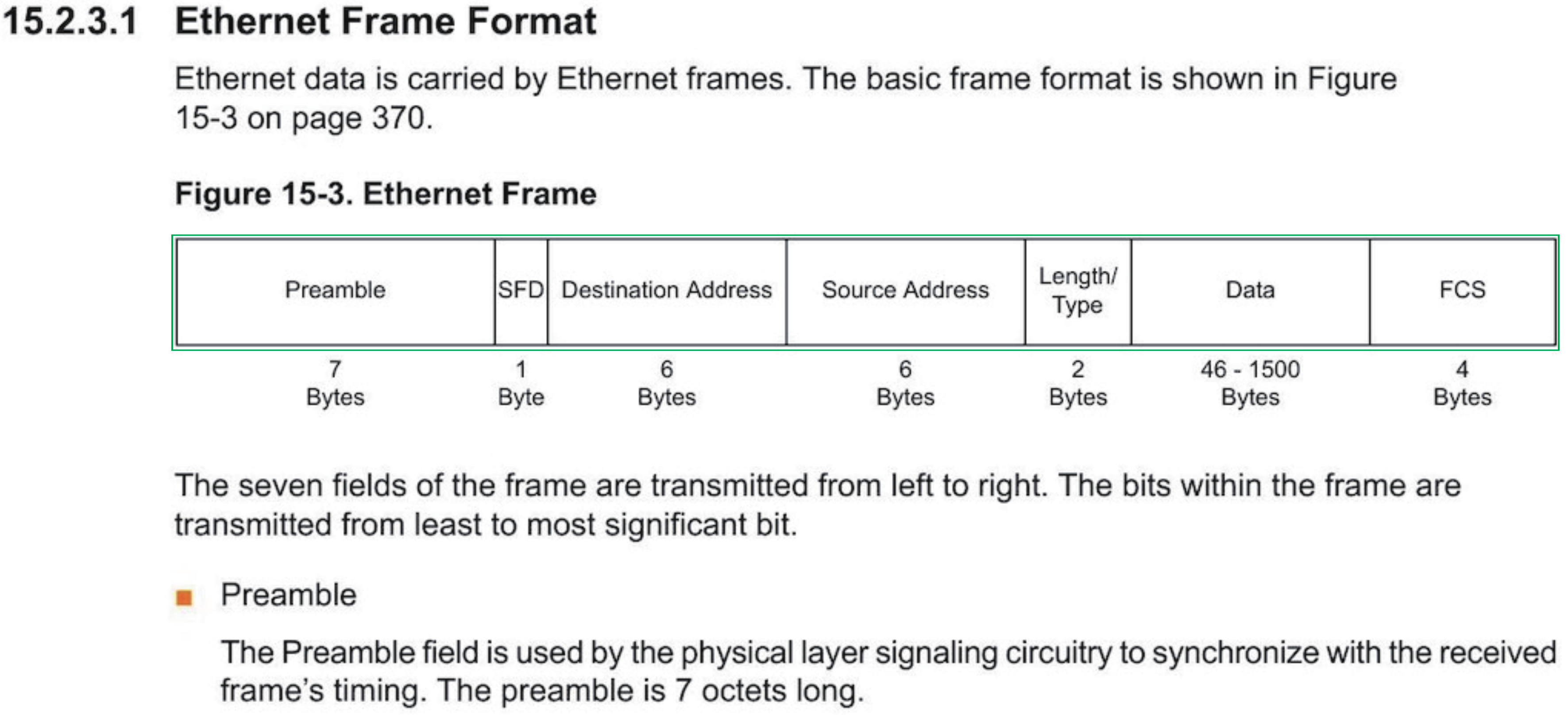}
  \caption{A sample of a single-row figure that is not annotated as a table. We highlight the single-row figure with a green box.}
  \label{fig:sample_of_single_row}
\end{center}
\end{figure}

\begin{figure}[htp]
\begin{center}
  \includegraphics[width=\columnwidth]{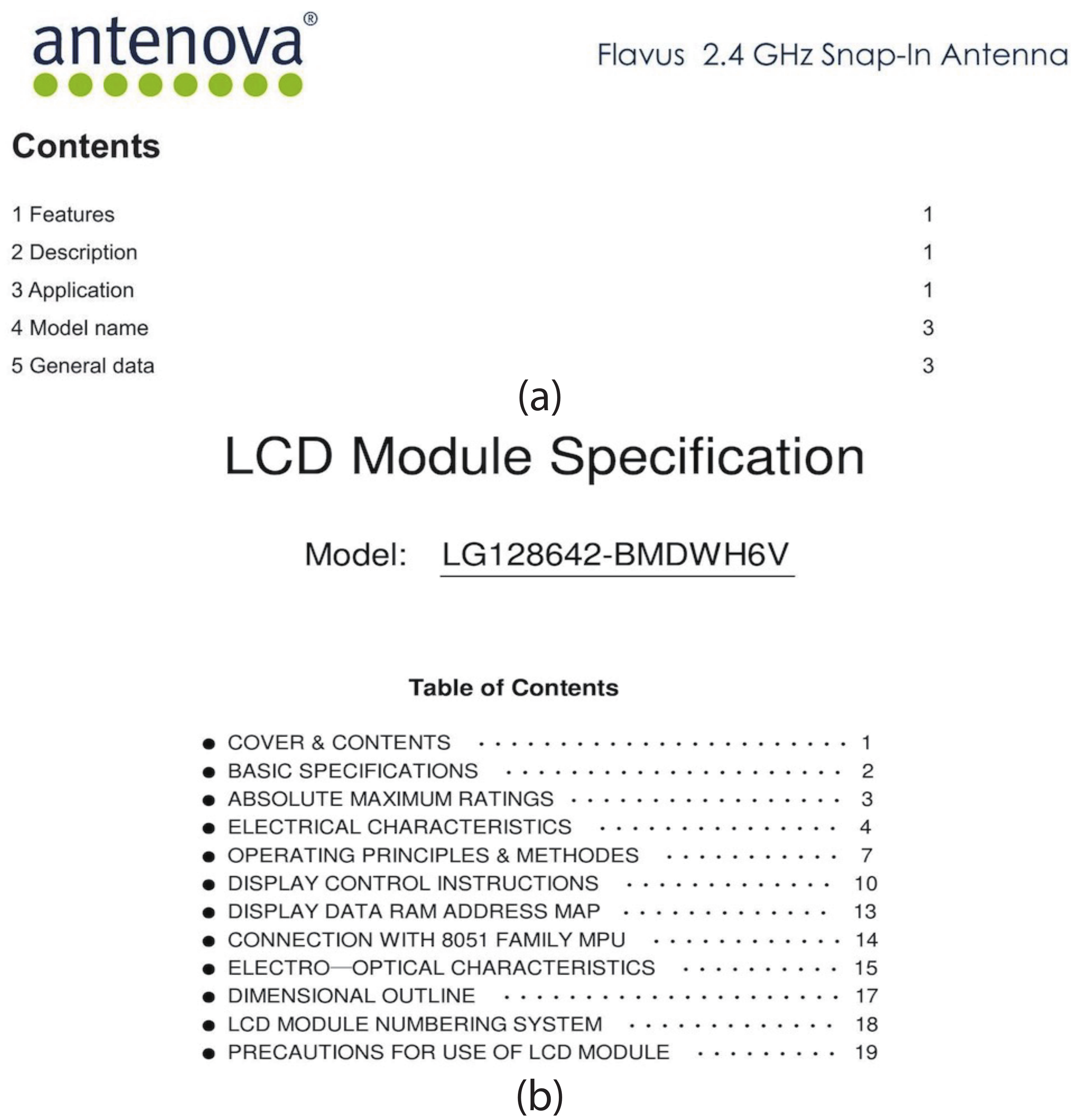}
  \caption{Two samples are not annotated as tables because they do not describe ICT commodities and are not useful for downstream tasks. Figure (a) is a sample whose appearance looks like a non-lined table but is not annotated as a table following our defined annotations rules. Figure (b) a sample is described as a Table but is not annotated as a Table following our defined annotations rules. }
  \label{fig:samles_are_not_table}
\end{center}
\end{figure}


Since many ambiguous cases exist in the ICT domain documents, we define the following rules to annotate tables. Firstly, a table must at least have two rows and two columns because a table should be a summary of critical information. We treat the ones with a single row or column as plain text or figures. Figure~\ref{fig:sample_of_single_row} shows an example of a single-row figure whose appearance resembles a table but should not be annotated as a table following this rule. Instead, it should be a figure. Secondly, tables should contain information describing the commodities because we want to extract information for domain-specific applications. Some information can be formatted like tables, such as the index page of the document, but not useful for the downstream tasks. Figure~\ref{fig:samles_are_not_table} shows two samples that are not annotated as tables because they are the index of content without containing information on any commodities. However, their appearances are similar to tables, making this dataset more challenging. Thirdly, table titles and table notes should not be included in the tables unless there are lines to have them as parts of a table because, in this study, we focus on the TD problem. Table titles and table notes should be treated as different components in a document, which is beyond the scope of this study.

Following these rules, tables in the proposed ICT-TD dataset can be grouped into four categories based on the content and the structure of tables: fully-lined tables, partially-lined tables, non-lined tables, and other unique tables. Figure~\ref{fig:table_samples} shows some samples of these different types of tables in the proposed ICT-TD dataset. Figure~\ref{fig:table_samples} (a) is a special table comprising many sub-tables. Since each of these sub-tables describes a parameter of the commodity, we treat the union of these sub-tables as a single special table.

\begin{table*}[ht!]
\caption{Statistics of the ICT-TD dataset and public datasets. Notably, the values in this table are the number of images, not the number of tables. * means the datasets are used to create the Open-Tables dataset.}
\centering
\begin{tabular}{  c | c | c | c | c | c }
\hline
\label{table:dataset_statistics}
Dataset & Type & Training & Testing & Validation & Data Source\\
\hline

TableBank~\cite{li2020tablebank} & Generated & $ 130,463 $ & $ 5,000 $ & $ 10,000 $ & Word and Latex   \\
PubLayNet~\cite{zhong2019publaynet} & Generated & $86,950$ & $ 3,772 $ & $ 3,950 $ &  PubMed  \\
\hline
ICDAR2013*~\cite{gobel2013icdar} & Manual & $ - $ & $ 150 $ & $ - $ & Governmental Documents  \\
ICDAR2017*~\cite{gao2017icdar2017} & Manual & $ 1600 $ & $ 817 $ & $ - $ & CiteSeer \\
ICDAR2019*~\cite{gao2019icdar} & Manual & $ 600 $ & $ 240 $ & $ - $ & Journals, Forms, Financial STMT \\
Marmot*~\cite{fang2012dataset} & Manual & $ 2000 $ & $ - $ & $ - $ &  E-book and CiteSeer \\
TNCR*~\cite{abdallah2022tncr} & Manual & $ 4634 $ & $ 987 $ & $ 1000 $ & Governmental Documents \\
IIIT-AR-13K~\cite{mondal2020iiit} & Manual & $ 9000 $ & $ 2000 $ & $ 2000 $ & Annual Reports \\
\hline
Open-Tables  & Manual & $8834$ & $ 1240 $ & $1000$ & Merged dataset \\
ICT-TD & Manual &  $ 4000 $ & $ 1000 $ & $ - $ & ICT PDF Documents \\

\hline
\end{tabular}
\end{table*}

\section{Experimental Results and Analysis}
\label{sec:experiments}
\subsection{Main Results}
\label{sec:main_results}
In this section, we conduct experiments to build baselines for the proposed dataset. We choose four state-of-the-art approaches including TableDet~\cite{fernandes2022tabledet}, DiffusionDet~\cite{chen2022diffusiondet}, Deformable-DETR~\cite{zhu2020deformable} and SparseR-CNN~\cite{sun2021sparse} as baseline models. TableDet is built on Cascade-RCNN~\cite{cai2018cascade} leveraging transfer learning and table-aware data augmentation to improve the performance for the TD problem further. Deformable-DETR is a typical transformer-based approach, and DiffusionDet introduces diffusion process~\cite{song2020denoising, ho2020denoising} to the object detection problem with random region proposals. SparseR-CNN is a typical method using learnable region proposals. Thus, our baseline models contain a two-stage model, a transformer-based model, a model with random proposals, and a model with learnable region proposals to cover the most popular object detectors. It is worth mentioning that we do not include one-stage detectors because one-stage detectors are usually not as good as the models we include here. We re-implemented TableDet with Detectron2~\cite{wu2019detectron2}, keeping the table-aware augmentation method. The implementation of Deformable-DETR can be found in detrex~\cite{ideacvr2022detrex}. DiffusionDet and SparseR-CNN have their official implementations. All these baseline models use ResNet50~\cite{he2016deep}, pre-trained on ImageNet~\cite{deng2009imagenet} as the training start point. Notably, the original design of TableDet uses pre-trained CascadeMaskR-CNN on COCO dataset~\cite{lin2014microsoft} as the initialization. We follow the default model parameter configurations of these benchmark models but tune some parameters regarding the training scheduling of DiffusionDet, Deformable-DETR, and SparseR-CNN because their default training scheduling parameters are tuned based on the COCO dataset. We summarize the scheduling parameters in Table~\ref{table:scheduling_parameters}. It is worth mentioning that all these benchmark models are built on Detection2. Thus, we follow the terms of Detectron2 in Table~\ref{table:scheduling_parameters}. 

\begin{table*}[ht!]
\caption{Key parameters of the benchmark models.}
\centering
\begin{tabular}{ c | c | c | c | c }
\hline
\label{table:scheduling_parameters}
Method & TableDet & DiffusionDet & Deformable-DETR & SparseR-CNN  \\
\hline
OPTIMIZER & SGD & AdamW & AdamW & AdamW \\
MAX\_ITER & 25,000 & 50,000 & 50,000 & 50,000  \\
MAX\_EPOCH & 100 & 200 & 200 & 200  \\
STEPS & - & 37,500 & 37,500 & 37,500 \\
SCHEDULER & - & MultiStepLR & MultiStepLR & MultiStepLR  \\
BASE\_LR & 1.0e-03 & 1.0e-05 & 1.0e-04 & 2.5e-05 \\
GAMMA & - & 0.1 & 0.1 & 0.1  \\
IMS\_PER\_BATCH & 16 & 16 & 16 & 16  \\
\hline
\end{tabular}
\end{table*}

We use precision, recall, and F1-score as the evaluation metrics. An IoU score is used as the threshold to determine whether a table is detected, which can be calculated by Equation~\ref{eq:iou_calculation}. Then, the True Positive is the number of predictions whose IoU scores to one of the ground truth bounding boxes are larger than an IoU threshold, and these corresponding ground truth bounding boxes are treated as being detected. Similarly, we can calculate the False Positive as the number of predictions whose IoU to all ground truths bounding boxes that are less than the IoU threshold, and the False Negative is the number of ground truth bounding boxes that are not detected. At last, the Precision, Recall and F1-score can be calculated by Equation~\ref{eq:precision_calculation},~\ref{eq:recall_calculation} and ~\ref{eq:f1_calculation}, respectively. 

As mentioned in section~\ref{sec:introduction}, the TD problem requires the detectors to maintain adequate precision and recall when the IoU threshold is high, and scores with larger IoU thresholds are more discriminate. Therefore, we follow the ICDAR2019 competition~\cite{gao2019icdar} to use weighted F1-score as the primary evaluation metric, which is defined as Equation~\ref{eq:weighted_f1_calculation}. We choose 80\%, 85\%, 90\%, and 95\% as the IoU thresholds instead of 60\%, 70\%, 80\%, and 90\% used in the ICDAR2019 competition. Besides, we also follow the experimental settings in study~\cite{abdallah2022tncr}, providing detailed experimental results regarding precision, recall, and F1-score with varying IoU thresholds from 50\% to 95\%. We include these detailed experimental results in ~\ref{sec:detailed_experimental_results}. 

\begin{figure*}[htp]
\begin{center}
  \includegraphics[width=\textwidth]{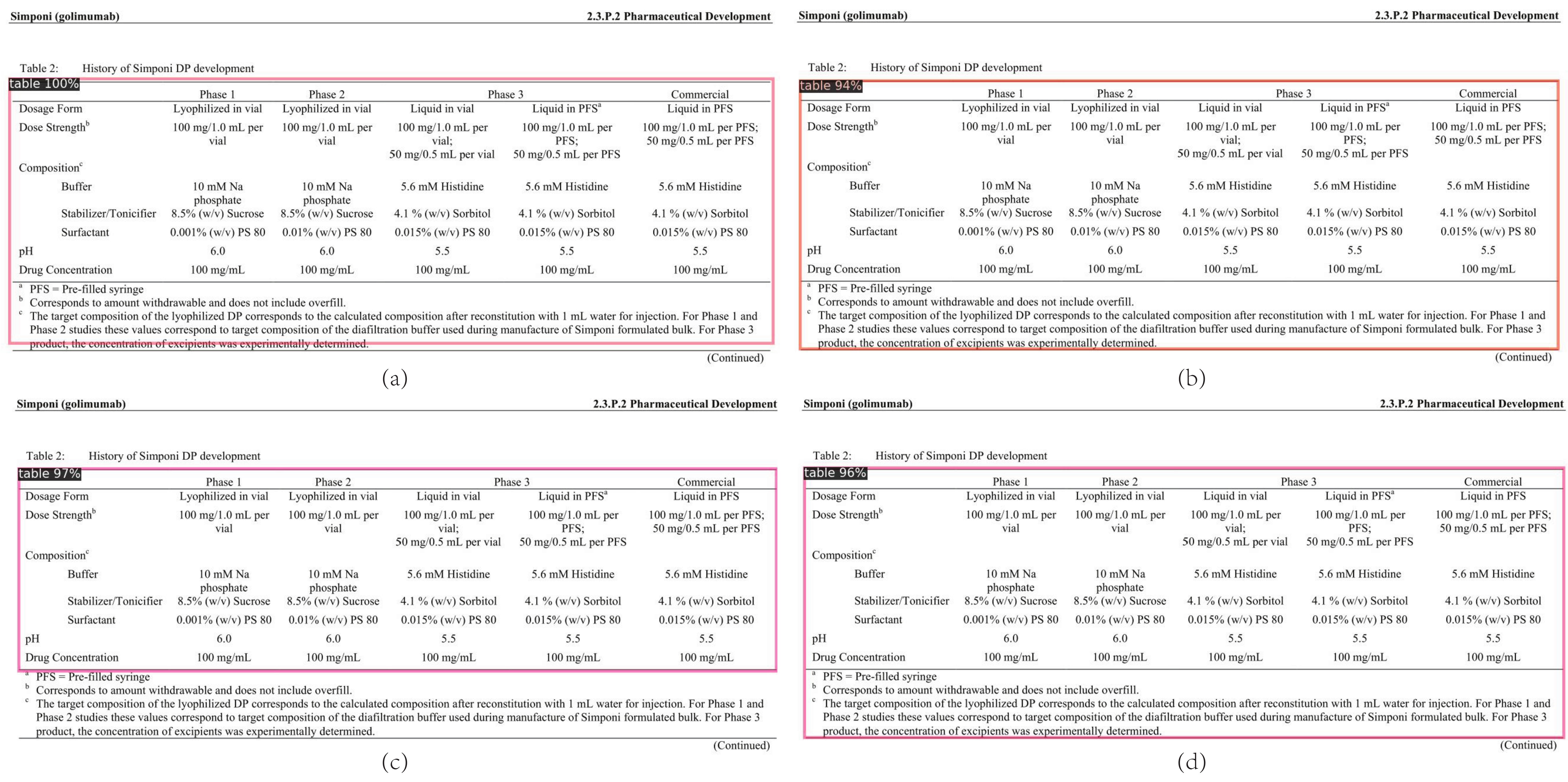}
  \caption{Prediction samples of the baseline models on the Open-Tables testing set. Figures (a) (b) (c) (d) are the results of TableDet, DiffusionDet, Deformable-DETR, and SparseR-CNN, respectively. The confidence scores in sub-figures are 100\%, 94\%, 97\% and 96\%, respectively.}
  \label{fig:prediction_samples_opentables1}
\end{center}
\end{figure*}

\begin{figure*}[htp]
\begin{center}
  \includegraphics[width=\textwidth]{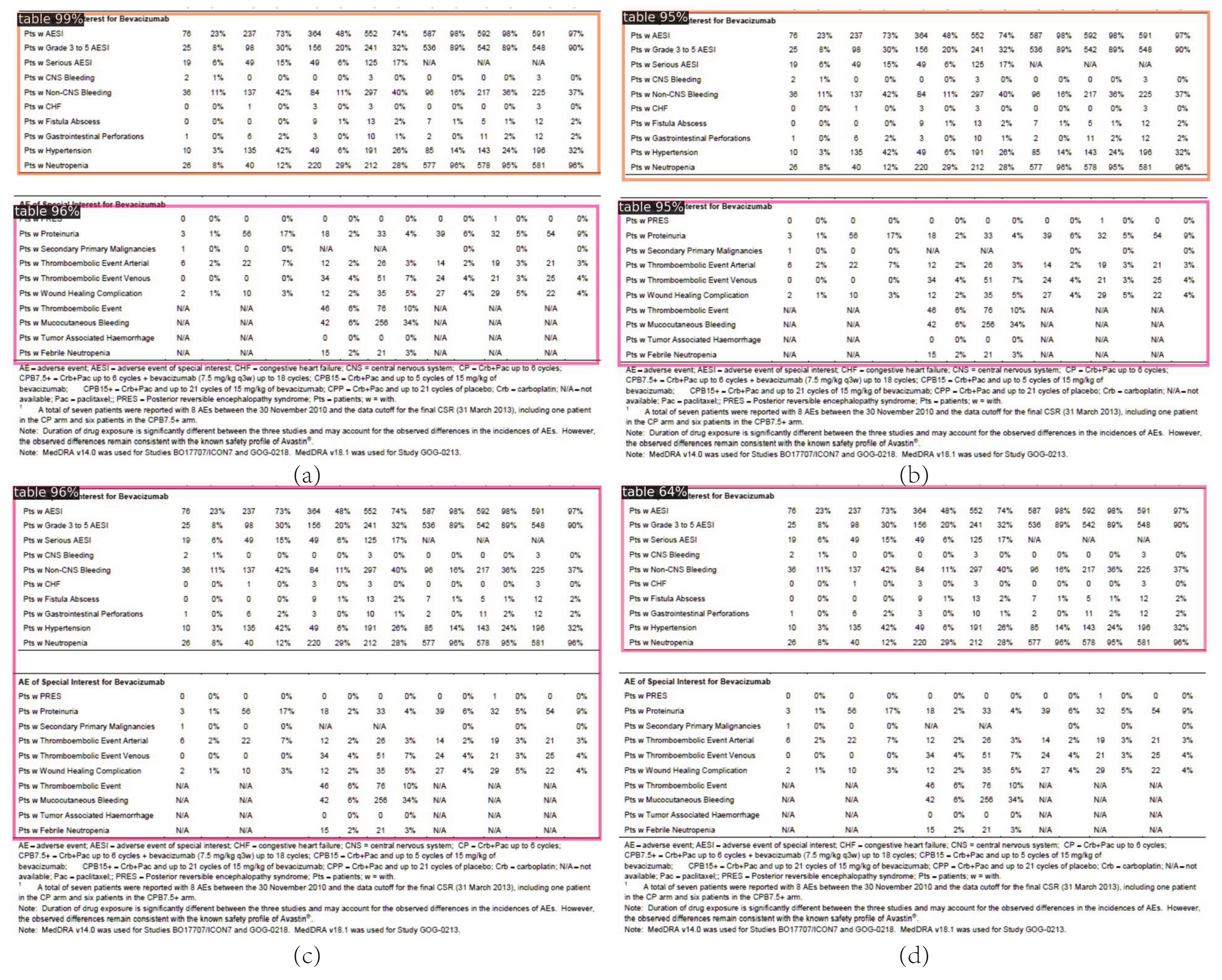}
  \caption{Prediction samples of the baseline models on the Open-Tables testing set. Figures (a) (b) (c) (d) are the results of TableDet, DiffusionDet, Deformable-DETR, and SparseR-CNN, respectively. The confidence scores in sub-figures are 99\%, 96\%, 95\%, 95\%, 96\% and 94\%, respectively.}
  \label{fig:prediction_samples_opentables2}
\end{center}
\end{figure*}

\begin{figure*}[htp]
\begin{center}
  \includegraphics[width=0\textwidth]{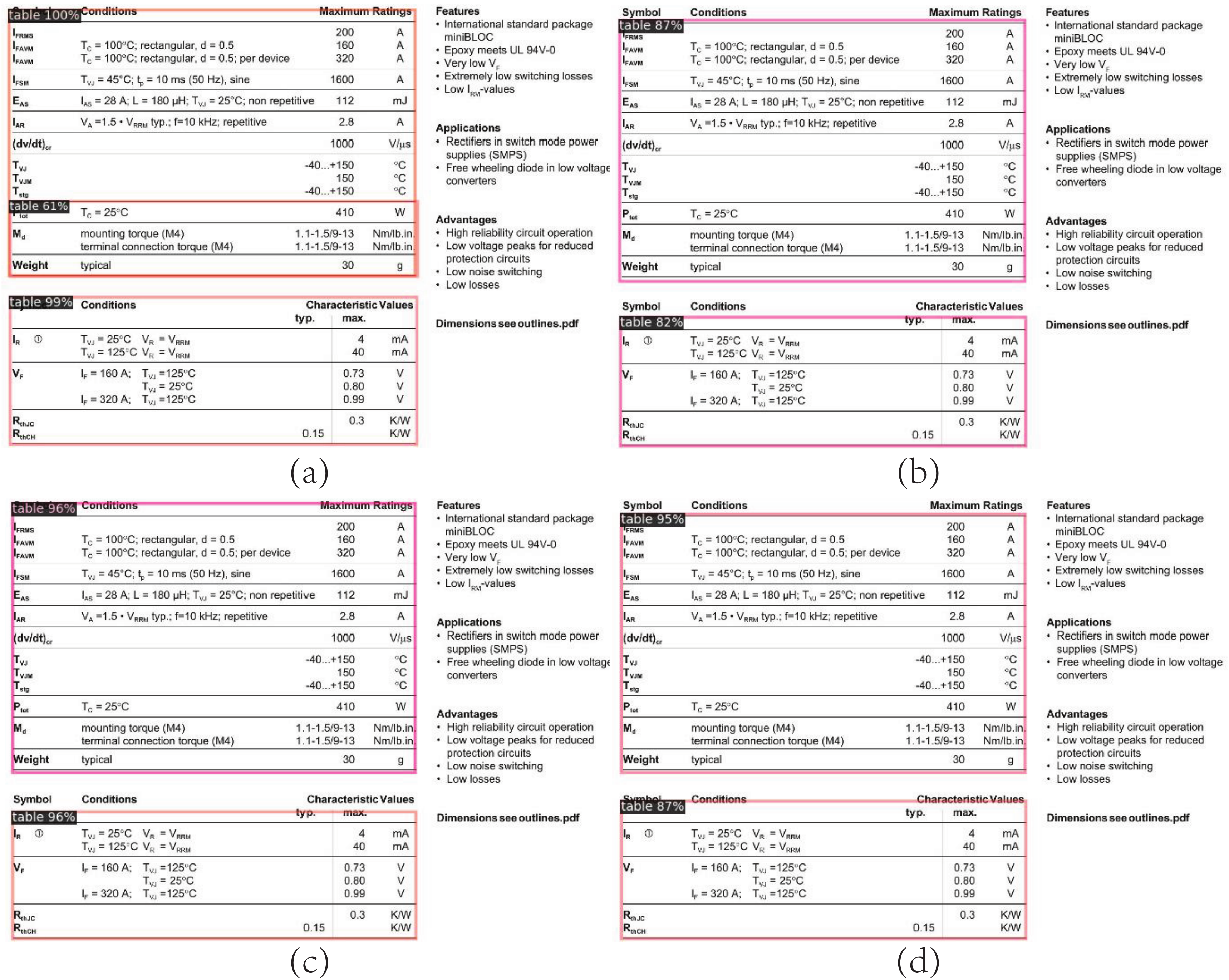}
  \caption{Prediction samples of the baseline models on the ICT-TD testing set. Figures (a) (b) (c) (d) are the results of TableDet, DiffusionDet, Deformable-DETR, and SparseR-CNN, respectively. The confidence scores in sub-figures are 100\%, 61\%, 99\%, 87\%, 82\%, 96\%, 96\%, 95\% and 87\%, respectively.}
  \label{fig:prediction_samples_icttd1}
\end{center}
\end{figure*}

\begin{figure*}[htp]
\begin{center}
  \includegraphics[width=\textwidth]{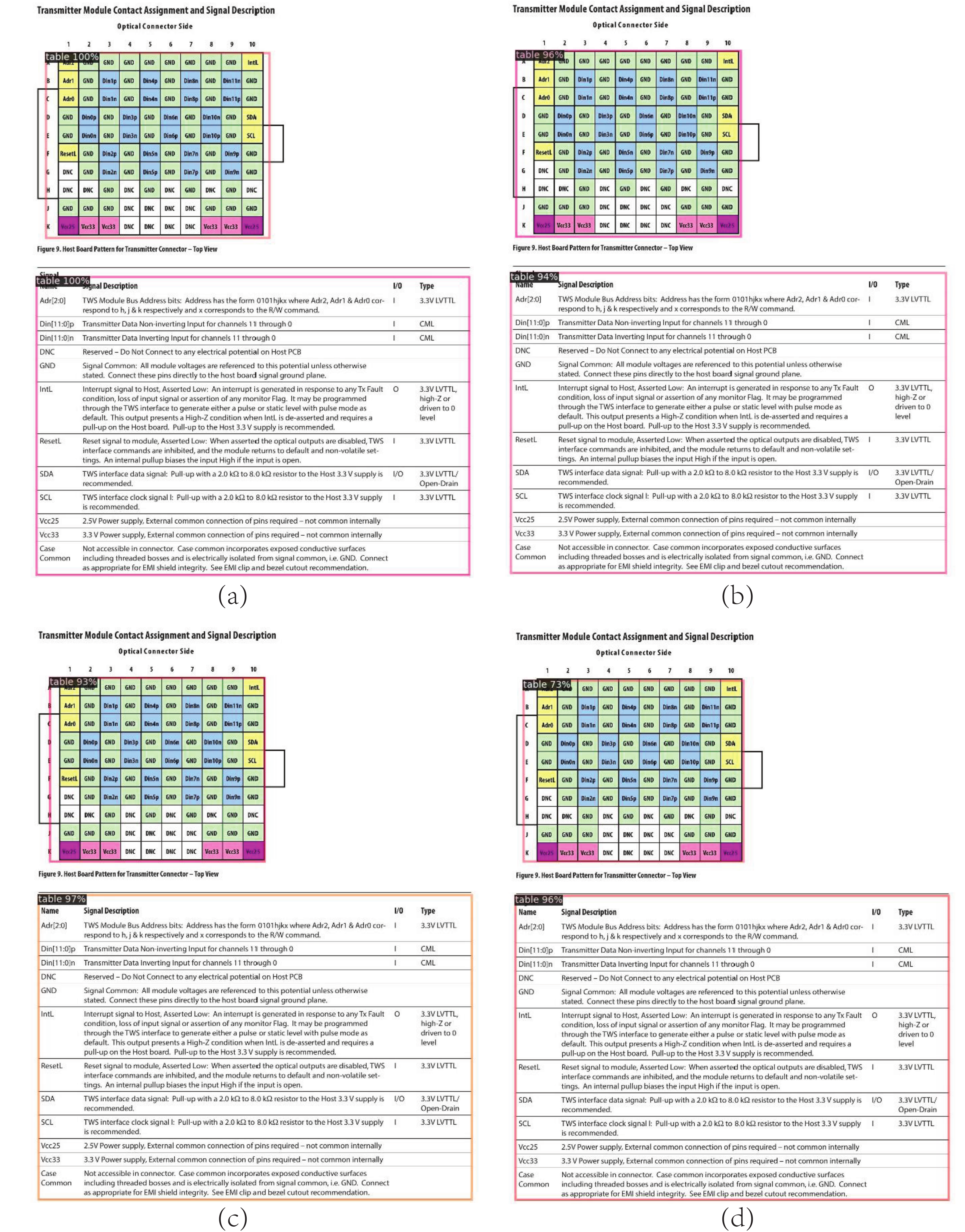}
  \caption{Prediction samples of the baseline models on the ICT-TD testing set. Figures (a) (b) (c) (d) are the results of TableDet, DiffusionDet, Deformable-DETR, and SparseR-CNN, respectively. The confidence scores in sub-figures are 100\%, 100\%, 96\%, 94\%, 93\%, 97\%, 73\%, and 96\%, respectively.}
  \label{fig:prediction_samples_icttd2}
\end{center}
\end{figure*}

Figures~\ref{fig:prediction_samples_opentables1} and ~\ref{fig:prediction_samples_opentables2} present some prediction samples of the baseline models. Sub-Figures (a) (b) (c) (d) are the results of TableDet, DiffusionDet, Deformable-DETR, and SparseR-CNN, respectively. For the table in Figure~\ref{fig:prediction_samples_opentables1}, its ground truth should contain the table explanation part because a bottom line bounds the explanation texts. However, the prediction box of TableDet is not large enough to cover all the explanation texts. Deformable-DETR doesn't treat explanation texts as part of the table, but its prediction box can fit other parts well. By contrast, DiffusionDet and SparseR-CNN can detect this table very well. For the table in Figure~\ref{fig:prediction_samples_opentables2}, TableDet and DiffusionDet can detect two tables successfully, even though their prediction boxes cannot fit the table precisely. By contrast, Deformable-DETR detects two tables as a single table, and SparseR-CNN missed the second table at the bottom. These samples show that the baseline models have different weaknesses in detecting tables from the proposed Open-Tables dataset, demonstrating that the Open-Tables dataset can be a useful source for TD studies. Similarly, we include several prediction samples on the ICT-TD dataset in Figures~\ref{fig:prediction_samples_icttd1} and ~\ref{fig:prediction_samples_icttd2}. The ideal boxes of tables in Figure~\ref{fig:prediction_samples_icttd1} should cover their table header cells that are not bounded by lines. However, only TabeDet can cover these header cells, but it detects the upper table as two tables. Figure~\ref{fig:prediction_samples_icttd2} shows a sample where all four baseline models recognize a figure as a table. It is worth mentioning that samples in Figures~\ref{fig:prediction_samples_icttd1} and ~\ref{fig:prediction_samples_icttd2} are domain-specific, making the ICT-TD dataset a useful source for ICT domain and cross-domain applications.

\begin{equation}
\label{eq:iou_calculation}
IoU = \frac{\textit{Overlap Area of two Boxes}}{\textit{Union Area of two Boxes}}
\end{equation}

\begin{equation}
\label{eq:precision_calculation}
\begin{aligned}
\text{Precision} = \frac{\text{True positive}}{\text{True positive} + \text{False positive}}
\end{aligned}
\end{equation}

\begin{equation}
\label{eq:recall_calculation}
\begin{aligned}
\text{Recall} = \frac{\text{True positive}}{\text{True positive} + \text{False negative}}
\end{aligned}
\end{equation}

\begin{equation}
\label{eq:f1_calculation}
\begin{aligned}
\text{F1-score} = 2*\frac{\text{Precision} * \text{Recall}}{\text{Precision} + \text{Recall}}
\end{aligned}
\end{equation}

\begin{equation}
\label{eq:weighted_f1_calculation}
\begin{aligned}
\text{Weighted Avg. F1-score} = \frac{\sum_{i=1}^{4} IoU_i \cdot F1@IoU_i}{\sum_{i=1}^4IoU_i}
\end{aligned}
\end{equation}

\begin{table*}[t]\caption{Experimental results on the ICT-TD dataset with F1-score. 4000 and 1000 are the number of samples.}
\centering
\begin{tabular}{ c  c  c  c  c  c  c  c }
 \toprule
\multicolumn{2}{c}{Dataset} & Model & \multicolumn{4}{c }{F1 under IoU thresholds} & Weight Avg. F1 \\ 
Training & Testing & & 80\% & 85\%  & 90\% & 95\% & \\ \midrule
\multirowcell{4}{ICT-TD \\ Training Set\\ (4000) } & \multirowcell{4}{ICT-TD \\ Testing Set\\ (1000) } & TableDet & 93.9 & 92.4 & 89.6 & 75.9 & 87.5  \\ 
& & DiffusionDet & 95.5 & 94.2 & 91.3 & 76.5 & 88.9 \\
& & Deformable-DETR & 95.1 & 93.8 & 91.6 & 82.1 & 90.3 \\
& & SparseR-CNN & 94.3 & 92.9 & 90.4 & 79.3 & 88.9 \\
\bottomrule
\end{tabular}
\label{table:icttd_results}
\end{table*}

\begin{table*}[t]
\centering
\caption{Experimental results on the Open-Tables dataset with F1-score. 8834 and 1240 are the number of samples. * means the models are trained with noisy samples.}
\begin{tabular}{ c c c c c c c c}
\toprule
\multicolumn{2}{c}{Dataset} &  Model & \multicolumn{4}{c}{F1 under IoU thresholds} & Weight Avg. F1 \\
Training & Testing &  & 80\% & 85\% & 90\% & 95\% &   \\ \midrule
\multirowcell{4}{Open-Tables \\ Training Set\\ (8834) } & \multirowcell{4}{Open-Tables \\ Testing Set\\ (1240) } & TableDet & 96.8 & 95.1 &  92.6 & 83.9 & 91.8  \\
& & DiffusionDet & 97.8 & 96.7 & 93.8 & 84.5 & 92.9 \\
& & Deformable-DETR & 96.7 & 95.3 & 93.7 & 87.6 & 93.1 \\
& & SparseR-CNN & 97.5 & 95.9 & 93.4 & 87.4 & 93.3 \\ \hline
\multirowcell{4}{Noisy \\ Open-Tables \\ Training Set\\ (8834) } & \multirowcell{4}{Open-Tables \\ Testing Set\\ (1240) } & TableDet* & 96.1 & 94.7 & 92.3 & 81.6 & 90.8  \\
& & DiffusionDet* & 97.4 & 95.8 & 93.4 & 84.4 & 92.4 \\
& & Deformable-DETR* & 96.3 & 94.9 & 91.8 & 84.9 & 91.7 \\
& & SparseR-CNN* & 96.9 & 95.6 & 93.2 & 85.7 & 92.6 \\ 
\bottomrule
\end{tabular}
\label{table:opentables_results}
\end{table*}

\subsection{Cross Domain Table Detection}
\label{sec:cross_domain_td}
In this section, we discuss the potential of using the proposal dataset in a cross-domain setting. As discussed in section~\ref{sec:ict_td_dataset}, ICDAR2013, ICDAR2017, ICDAR2019, Marmot, and TNCR are also manually annotated datasets with high-quality annotations, and their data sources are academic publications and open governmental documents. Therefore, we merge these datasets, resulting in a new dataset, termed Open-Tables, which contains 8834 training samples, 2282 testing samples, and 1000 validation samples. It is worth mentioning that TNCR has five different groups of tables, as discussed in section~\ref{sec:related_work}. We simply merged all these groups into a single group as tables. After the cleaning tasks discussed in section~\ref{sec:open_tables}, two cross-domain settings are used to build the benchmarks. First, we use ICT-TD's training set to train the detection baseline models and Open-Tables' test set to evaluate the model's performance. The experimental results of this setting are shown in Table~\ref{table:cross_domain_icttd_to_aca}. In the second setting, the training set of the Open-Tables dataset and the testing set of the ICT-TD dataset are used to build the benchmarks, and the results are shown in Table~\ref{table:crossdomain_results_opentables_to_icttd}. It is worth mentioning that we use the same evaluation metrics as section ~\ref{sec:main_results}, and the * in Table~\ref{table:crossdomain_results_opentables_to_icttd} means the models are trained with the noisy version of Open-Tables dataset.

The experimental results show that Deformable-DETR, which performs best for the ICT-TD dataset, also has the best generalization capacity in the cross-domain setting. However, the cross-domain setting is much more challenging, and all the benchmark models' performance degrades by a large margin compared with results in Table~\ref{table:icttd_results} and Table~\ref{table:opentables_results}.

\begin{table*}[t]
\caption{Experimental results with F1-score in the cross domain setting. 4000 and 1240 are the number of samples.}
\centering
\begin{tabular}{ c c c c c c c c}
 \toprule
\multicolumn{2}{c}{Dataset} & Model & \multicolumn{4}{c }{F1 under IoU thresholds} & Weight Avg. F1 \\ 
Training & Testing & & 80\% & 85\%  & 90\% & 95\% & \\ \midrule
\multirowcell{4}{ICT-TD \\ Training Set \\ (4000) } & \multirowcell{4}{Open-Tables \\ Testing Set\\ (1240) } & TableDet & 77.8 & 74.2 &  68.6 & 52.1 & 67.6  \\ 
&&DiffusionDet & 85.3 & 80.8 & 72.6 & 53.7 & 72.4 \\
&&Deformable-DETR & 85.2 & 81.4 & 75.9 & 62.6 & 75.8 \\
&&SparseR-CNN & 80.0 & 76.3 & 69.7 & 55.8 & 69.9 \\
\bottomrule
\end{tabular}
\label{table:cross_domain_icttd_to_aca}
\end{table*}

\begin{table*}[t]
\caption{Experimental results with F1-score in the cross domain setting. 8834 and 1000 are the number of training samples. * means the models are trained with noisy samples.}
\centering
\begin{tabular}{ c c c c c c c c}
 \toprule
\multicolumn{2}{c}{Dataset} & Model & \multicolumn{4}{c }{F1 under IoU thresholds} & Weight Avg. F1 \\ 
Training & Testing & & 80\% & 85\%  & 90\% & 95\% & \\ \midrule
\multirowcell{4}{Open-Tables \\ Training Set\\ (8834) } & \multirowcell{4}{ICT-TD \\ Testing Set\\ (1000) }&TableDet & 83.1 & 80.1 & 76.1 & 65.0 & 75.7  \\ 
& &DiffusionDet & 87.9 & 86.1 & 81.6 & 67.0 & 80.2 \\
& &Deformable-DETR & 84.1 & 82.2 & 79.7 & 70.1 & 78.7 \\
& &SparseR-CNN & 84.2 & 81.9 & 78.5 & 67.8 & 77.7 \\  \hline
\multirowcell{4}{Noisy \\ Open-Tables \\ Training Set\\ (8834) }& \multirowcell{4}{ICT-TD \\ Testing Set\\ (1000) } &TableDet* & 81.0 & 78.2 & 73.8 & 61.1 & 73.1  \\
& &DiffusionDet* & 86.2 & 84.0 & 79.8 & 66.3 & 78.6 \\
& &Deformable-DETR* & 81.9 & 80.4 & 77.9 & 67.7 & 76.7 \\
& &SparseR-CNN* & 85.0 & 82.5 & 78.4 & 64.2 & 77.1 \\
\bottomrule
\end{tabular}
\label{table:crossdomain_results_opentables_to_icttd}
\end{table*}


\subsection{The Impact of Noise in Open-Tables Dataset}
We conducted extra experiments to discuss the impact of noise in the Open-Tables dataset. In the experiments, the training set of Open-Tables with noise is used to train the benchmark models, and the testing set of the ICT-TD dataset is used to evaluate the model performance. As shown in Table~\ref{table:crossdomain_results_opentables_to_icttd}, the cleaned version of the Open-Tables dataset can improve the performance of all models, especially when the IoU threshold is above 80\%. The experimental results verify the necessity of noise cleaning and label alignment when we create the Open-Tables dataset. Besides, we also evaluate these models with the cleaned Open-Tables testing set, and the experimental results are given in Table~\ref{table:opentables_results}. Similar to the results shown in Figure~\ref{table:crossdomain_results_opentables_to_icttd}, the models trained with the cleaned Open-Tables training set perform better than their counterparts trained with the noisy Open-Tables training set. It is worth mentioning that the Open-Tables testing set is created by merging the cleaned testing set of ICDAR2013, ICDAR2017, ICDAR2019, and TNCR datasets, as shown in Table~\ref{table:dataset_statistics}. Therefore, the results shown in Table~\ref{table:opentables_results} also reflect that these existing datasets can benefit from our proposed Open-Tables dataset.

\subsection{Potential Applications}
In this section, we discuss the potential applications of the proposed two datasets. As mentioned earlier, the ICT-TD dataset is created using real documents from the ICT domain. The cross-domain setting that uses Open-Tables' training set to train the models and then test them on the ICT-TD testing set does not perform well, as shown in Table~\ref{table:crossdomain_results_opentables_to_icttd}. By contrast, as shown in Table~\ref{table:icttd_results} ~\ref{table:icttd_detailed_results}, the models trained with the ICT-TD training set can achieve much better results when tested on the ICT-TD testing set. Therefore, we argue that the proposed ICT-TD dataset can be used to train and evaluate ICT domain-specific models and applied to the ICT supply chain optimization problems~\cite{6693469} as part of the information processing step. Besides, the proposed ICT-TD dataset can also enrich the data sources of public datasets and be used to evaluate models' generalization ability in cross-domain settings. On the other hand, the Open-Tables dataset focuses on addressing the noise issues of existing public datasets. As shown in Table~\ref{table:crossdomain_results_opentables_to_icttd}, a cleaned version of Open-Tables' training set improves the models' generalization ability in the cross-domain settings. Furthermore, the Open-Tables' testing set is created by merging the cleaned testing sets of ICDAR2013, ICDAR2017, ICDAR2019, and TNCR datasets, which means that it can provide more reliable evaluation results.

\section{Conclusion}
\label{sec:conclusion}
In this paper, we revisit some popular datasets with high-quality annotations but different annotation definitions, clean the noisy samples, and align the annotations of these datasets to form a larger, high-quality dataset termed Open-Tables. Since the data sources of popular datasets are very limited, we propose a new dataset termed ICT-TD using the datasheets from the ICT domain. Our proposed ICT-TD dataset contains many domain-specific samples that hardly appear in other open datasets to make it useful in cross-domain settings. The revisited Open-Tables dataset is consistent and larger, making it more reliable to evaluate the model performance. These two datasets can be more reliable benchmarks to build reliable TD applications that should avoid losing any information in the tables and alleviate the side effects of noisy samples to the model evaluation. At last, we build strong baselines using state-of-the-art object detection models for the ICT-TD dataset and a cross-domain setting. The experimental results show that cross-domain settings are more challenging for the TD problem.

Most existing studies for the TD problem use object detection evaluation metrics that need an IoU threshold. However, these evaluation metrics are indirect to the actual performance of extracting information from tables. For instance, a larger prediction box that can cover all the information of the target table but has a lower IoU score is preferable to the box with a higher IoU score but can lose some information from the target table. Therefore, evaluating models with other metrics can be a good direction for further work to compensate for the drawback of using IoU score based metrics.

\bibliographystyle{sn-aps}
\bibliography{sn-bibliography}

\section{Appendix}
\subsection{Detailed experimental results}
\label{sec:detailed_experimental_results}
In this section, we list the detailed experimental results. More specifically, Table~\ref{table:icttd_detailed_results} and Table~\ref{table:opentables_detailed_results} show the results on the ICT-TD and Open-Tables datasets, respectively. Table~\ref{table:opentables_cross_domain_detailed_results} and Table~\ref{table:icttd_crossdomain_detailed_results} are the results of two cross-domain settings, namely using the training set of ICT-TD dataset and the testing set of Open-Tables dataset as the first setting, and using the training set of Open-Tables dataset and the testing set of ICT-TD dataset as the second setting. 

\begin{table*}[t]
\caption{Detailed Experimental results on the ICT-TD dataset.}
\centering
\begin{tabular}{ c c c c c c c c c c c c c c}
 \toprule
\label{table:icttd_detailed_results}
\multirow{2}{*}{Method} & \multirow{2}{*}{Metric} & \multicolumn{11}{{c}}{IoU}  \\ \cmidrule{3-13}
& &50\% &55\% &60\% &65\% &70\% &75\% &80\% &85\% &90\% &95\% & 50\%:95\% \\ \midrule
TableDet  & Precision & 97.0 & 96.1 & 96.1 & 95.9 & 94.9 & 94.0 & 93.0 & 91.6 & 88.5 & 73.2 & 92.0 \\ 
 & Recall & 98.0 & 97.6 & 97.5 & 97.2 & 96.5 & 95.8 & 94.8 & 93.3 & 90.6 & 78.9 & 94.0\\ 
 &  F1 & 97.5 & 96.8 & 96.8 & 96.5 & 95.7 & 94.9 & 93.9 & 92.4 & 89.5 & 75.9 & 93.0 \\ 

\hline
DiffusionDet  & Precision & 96.9 & 96.7 & 96.5 & 96.2 & 95.2 & 94.7 & 93.8 & 92.4 & 89.4  & 73.8 & 92.6 \\ 
 &   Recall & 99.2 & 99.0 & 98.9 & 98.8 & 98.4 & 97.9 & 97.2 & 96.1 & 93.2 & 79.4 & 95.8 \\ 
 &  F1 & 98.0 & 97.8 & 97.7 & 97.5 & 96.8 & 96.3 & 95.5 & 94.2 & 91.3 & 76.5 & 94.2 \\ 

\hline
Deformable-  & Precision & 97.0 & 96.7 & 96.4 & 96.0 & 95.1 & 94.4 & 93.8 & 92.4 & 90.0 & 79.4 & 93.1\\ 
 DETR& Recall & 98.9 & 98.8 & 98.6 & 98.4 & 97.4 & 96.9 & 96.5 & 95.2 & 93.2 & 84.9 & 95.9\\ 
 &  F1 & 97.9 & 97.7 & 97.5 & 97.2 & 96.2 & 95.6 & 95.1 & 93.8 & 91.6 & 82.1 & 94.5\\ 
 \hline
SparseR-CNN & Precision & 95.8 & 95.5 & 95.2 & 95.0 & 94.3 & 93.6 & 92.6 & 91.1 & 88.4 & 75.6 & 91.7\\ 
 &  Recall & 98.7 & 98.4 & 98.2 & 98.1 & 97.5 & 97.2 & 96.1 & 94.7 & 92.5 & 83.4 & 95.5 \\ 
 &  F1 & 97.2 & 96.9 & 96.7 & 96.5 & 95.7 & 95.4 & 94.3 & 92.9 & 90.4 & 79.3 & 93.6 \\ 
 \hline
 \bottomrule 
\end{tabular}
\end{table*}

\begin{table*}[t]
\caption{Detailed Experimental results on the Open-Tables dataset.}
\centering
\begin{tabular}{ c c c c c c c c c c c c c c}
 \toprule
\label{table:opentables_detailed_results}
\multirow{2}{*}{Method} & \multirow{2}{*}{Metric} & \multicolumn{11}{{c}}{IoU}  \\ \cmidrule{3-13}
& &50\% &55\% &60\% &65\% &70\% &75\% &80\% &85\% &90\% &95\% & 50\%:95\% \\ \midrule
TableDet  & Precision & 98.7 & 97.9 & 97.9 & 97.8 & 97.8 & 96.7 & 96.6 & 94.3 & 91.7 & 81.8 & 95.1 \\ 
 & Recall & 99.1 & 99.0 & 98.9 & 98.5 & 98.4 & 97.7 & 97.1 & 95.9 & 93.5 & 86.1 & 96.4\\ 
 &  F1 & 98.9 & 98.4 & 98.4 & 98.1 & 98.1 & 97.2 & 96.8 & 95.1 & 92.6 & 83.9 & 95.7 \\ 

\hline
DiffusionDet  & Precision & 98.3 & 98.2 & 98.0 & 97.9 & 97.8 & 97.2 & 96.8 & 95.3 & 92.0  & 81.8 & 95.3 \\ 
 &   Recall & 99.8 & 99.7 & 99.7 & 99.7 & 99.6 & 99.4 & 98.9 & 98.1 & 95.6 & 87.4 & 97.8 \\ 
 &  F1 & 99.1 & 99.0 & 98.8 & 98.8 & 98.7 & 98.3 & 97.8 & 96.7 & 93.8 & 84.5 & 96.6 \\ 

\hline
Deformable-  & Precision & 97.8 & 97.7 & 97.7 & 97.5 & 97.0 & 96.5 & 95.8 & 94.1 & 92.0 & 85.2 & 95.1\\ 
DETR & Recall & 99.4 & 99.3 & 99.2 & 99.0 & 98.6 & 98.0 & 97.6 & 96.6 & 95.3 & 90.3 & 97.3\\ 
 &  F1 & 98.6 & 98.5 & 98.4 & 98.3 & 97.8 & 97.2 & 96.7 & 95.3 & 93.7 & 87.6 & 96.2\\ 
 \hline
SparseR-CNN & Precision & 98.4 & 98.2 & 98.1 & 97.9 & 97.7 & 97.0 & 96.3 & 94.2 & 91.4 & 84.5 & 95.4\\ 
 &  Recall & 99.7 & 99.7 & 99.7 & 99.7 & 99.5 & 99.1 & 98.7 & 97.7 & 95.5 & 90.5 & 98.0 \\ 
 &  F1 & 99.1 & 99.0 & 98.9 & 98.8 & 98.6 & 98.0 & 97.5 & 95.9 & 93.4 & 87.4 & 96.7 \\ 
 \hline
 \bottomrule 
\end{tabular}
\end{table*}

\begin{table*}[t]
\caption{Detailed Experimental results in the cross domain setting. The training set is from ICT-TD and the testing set is from Open-Tables.}
\centering
\begin{tabular}{ c c c c c c c c c c c c c c}
 \toprule
\label{table:opentables_cross_domain_detailed_results}
\multirow{2}{*}{Method} & \multirow{2}{*}{Metric} & \multicolumn{11}{{c}}{IoU}  \\ \cmidrule{3-13}
& &50\% &55\% &60\% &65\% &70\% &75\% &80\% &85\% &90\% &95\% & 50\%:95\% \\ \midrule
TableDet & Precision & 88.4 & 87.2 & 86.0 & 84.9 & 82.8 & 79.4 & 75.7 & 72.0 & 66.4 & 48.6 & 77.1 \\ 
 & Recall & 91.9 & 90.4 & 89.4 & 88.3 & 86.6 & 83.7 & 80.1 & 76.6 & 71.0 & 56.0 & 81.4 \\ 
 &  F1 & 90.1 & 88.8 & 87.7 & 86.6 & 84.7 & 81.5 & 77.8 & 74.2 &  68.6 & 52.1 & 79.2 \\ 
 
\hline
DiffusionDet  & Precision & 86.2 & 85.5 & 84.8 & 83.7 & 82.3 & 80.5 & 77.5 & 72.7 & 65.3  & 48.5 & 76.7 \\ 
 &   Recall & 98.8 & 98.4 & 98.3 & 97.9 & 97.4 & 96.6 & 94.9 & 91.0 & 81.8 & 60.1 & 91.5 \\
 &  F1 & 92.1 & 91.5 & 91.1 & 90.2 & 89.2 & 87.8 & 85.3 & 80.8 & 72.6 & 53.7 & 83.4 \\ 
 
\hline
Deformable-  & Precision & 91.2 & 90.4 & 89.5 & 88.3 & 86.6 & 84.1 & 80.9 & 76.9 & 71.3 & 57.3 & 81.6\\ 
DETR & Recall & 97.3 & 96.7 & 96.0 & 95.4 & 94.1 & 93.0 & 90.0 & 86.5 & 81.2 & 69.0 & 89.9\\ 
 &  F1 & 94.2 & 93.4 & 92.6 & 91.7 & 90.2 & 88.3 & 85.2 & 81.4 & 75.9 & 62.6 & 85.5\\
 
 \hline
SparseR-CNN & Precision & 85.6 & 84.6 & 83.4 & 81.9 & 80.3 & 78.0 & 74.5 & 70.8 & 64.7 & 50.9 & 75.5 \\ 
 &  Recall & 97.2 & 96.3 & 95.2 & 93.9 & 92.1 & 89.9 & 86.4 & 82.7 & 75.5 & 61.8 & 87.1 \\
 &  F1 & 91.0 & 90.1 & 88.9 & 87.5 & 85.8 & 83.5 & 80.0 & 76.3 & 69.7 & 55.8 & 80.9 \\ 
 \hline
 \bottomrule 
\end{tabular}
\end{table*}

\begin{table*}[t]
\caption{Detailed experimental results in the cross domain setting. The training set is from Open-Tables and the testing set is from ICT-TD .}
\centering
\begin{tabular}{ c c c c c c c c c c c c c c}
 \toprule
\label{table:icttd_crossdomain_detailed_results}
\multirow{2}{*}{Method} & \multirow{2}{*}{Metric} & \multicolumn{11}{{c}}{IoU}  \\ \cmidrule{3-13}
& &50\% &55\% &60\% &65\% &70\% &75\% &80\% &85\% &90\% &95\% & 50\%:95\% \\ \midrule
TableDet  & Precision & 86.5 & 86.2 & 85.2 & 85.0 & 84.0 & 83.6 & 81.8 & 78.7 & 74.4 & 62.3 & 80.8 \\ 
 & Recall & 88.9 & 88.4 & 87.7 & 87.3 & 86.7 & 86.1 & 84.5 & 81.5 & 77.9 & 67.9 & 83.7 \\ 
 &  F1 & 87.7 & 87.3 & 86.4 & 86.1 & 85.3 & 84.8 & 83.1 & 80.1 & 76.1 & 65.0 & 82.2 \\ 

\hline
DiffusionDet  & Precision & 88.1 & 87.4 & 86.9 & 86.5 & 86.0 & 85.1 & 83.7 & 82.0 & 77.6 & 63.8 & 82.7 \\ 
 & Recall & 95.8 & 95.3 & 95.2 & 94.8 & 94.3 & 93.9 & 92.5 & 90.7 & 86.2 & 70.6 & 90.9\\ 
 &  F1 & 91.8 & 91.2 & 90.9 & 90.5 & 90.0 & 89.3 & 87.9 & 86.1 & 81.6 & 67.0 & 86.6 \\ 

\hline
Deformable- & Precision & 88.4 & 87.7 & 87.0 & 85.7 & 83.5 & 81.9 & 80.2 & 78.7 & 76.3 & 66.3 & 81.6 \\ 
DETR & Recall & 94.5 & 94.0 & 93.4 & 92.8 & 91.1 & 89.9 & 88.4 & 86.2 & 83.3 & 74.4 & 88.8\\ 
 &  F1 & 91.3 & 90.7 & 90.1 & 89.1 & 87.1 & 85.7 & 84.1 & 82.2 & 79.7 & 70.1 & 85.0 \\ 
 \hline
SparseR-CNN & Precision & 84.5 & 83.9 & 83.4 & 82.9 & 81.8 & 80.8 & 79.7 & 77.4 & 73.9 & 63.4 & 79.2 \\ 
 & Recall & 94.9 & 94.4 & 94.1 & 93.5 & 92.5 & 91.2 & 89.4 & 86.9 & 83.6 & 72.8 & 89.3\\ 
 &  F1 & 89.4 & 88.8 & 88.4 & 87.9 & 86.8 & 85.7 & 84.2 & 81.9 & 78.5 & 67.8 & 83.9 \\ 
 \hline
 \bottomrule 
\end{tabular}
\end{table*}


\end{document}